\documentclass[aip,reprint,showpacs,twocolumn,superscriptaddress,floatfix]{revtex4-1}
\usepackage{epsfig}
\usepackage{amsmath}
\usepackage{amssymb}
\usepackage{amsfonts}
\usepackage{mathptmx}
\usepackage{dcolumn}
\usepackage{eucal}
\usepackage{bm}
\usepackage{color}
\usepackage[colorlinks,linkcolor=blue,citecolor=blue]{hyperref}
\usepackage{epstopdf}

\usepackage{graphicx}

\usepackage{natbib}

\newcommand{\tmpnote}[1]%
   {\begingroup{\color{blue}\it (FIXME: #1)}\endgroup}

\begin{document}
\title{ Proposal for a phase-coherent thermoelectric transistor}
\author{F. Giazotto}
\email{giazotto@sns.it}
\affiliation{NEST, Instituto Nanoscienze-CNR and Scuola Normale Superiore, I-56127 Pisa, Italy}
\author{J. W. A. Robinson}
\email{jjr33@cam.ac.uk}

\affiliation{\mbox{Department of Materials Science and Metallurgy, University of Cambridge, 27 Charles Babbage Road, Cambridge CB3 0FS, UK}}
\author{J. S. Moodera}
\affiliation{Department of Physics and Francis Bitter Magnet Lab, Massachusetts Institute of Technology, Cambridge, Massachusetts 02139, USA}
\author{F. S. Bergeret}
\email{sebastian\_bergeret@ehu.es}
\affiliation{\mbox{Centro de F\'{i}sica de Materiales (CFM-MPC), Centro
Mixto CSIC-UPV/EHU, Manuel de Lardizabal 4, E-20018 San
Sebasti\'{a}n, Spain}}
\affiliation{Donostia International Physics Center (DIPC), Manuel
de Lardizabal 5, E-20018 San Sebasti\'{a}n, Spain}

\begin{abstract}

Identifying  materials and devices which offer efficient thermoelectric effects at low temperature is a major obstacle for the development of thermal management strategies for low-temperature electronic systems. Superconductors cannot offer a solution since their near perfect electron-hole symmetry leads to a negligible  thermoelectric response; however, here we demonstrate theoretically a superconducting thermoelectric transistor which offers unparalleled figures of merit of up to $\sim 45$ and Seebeck coefficients as large as a few mV/K at sub-Kelvin temperatures. The device is also phase-tunable meaning its thermoelectric response for power generation can be precisely controlled with a small magnetic field.
Our concept is based on a superconductor-normal metal-superconductor interferometer in which the normal metal weak-link is tunnel coupled to a ferromagnetic insulator and a Zeeman split superconductor . Upon  application of an external magnetic flux,  the interferometer enables phase-coherent manipulation of thermoelectric properties whilst offering efficiencies which approach the Carnot limit. 

\end{abstract}

\maketitle

It is known that electron-hole symmetry breaking is essential for a material to posses a finite thermoelectric figure of merit\cite{Ashcroft,Mahan}. In principle,  conventional superconductors have a near perfect symmetric spectrum and therefore are not suitable for thermoelectric devices. However, if the density of states is spin-split by a Zeeman field a superconductor-ferromagnet hybrid device can provide a thermoelectric effect \cite{machon2013,ozaeta2014,machon2013n2} with a figure of merit close to 1.  Here we propose a multifunctional phase-coherent superconducting  
transistor in which the thermoelectric efficiency is tunable through an externally applied magnetic flux.  A giant Seebeck coefficient of several mV/K and a figure of merit close to $\sim 45$  is predicted for realistic materials parameters and materials combinations.

\begin{figure*}[th!]
\begin{center}
\includegraphics[width=0.65\textwidth]{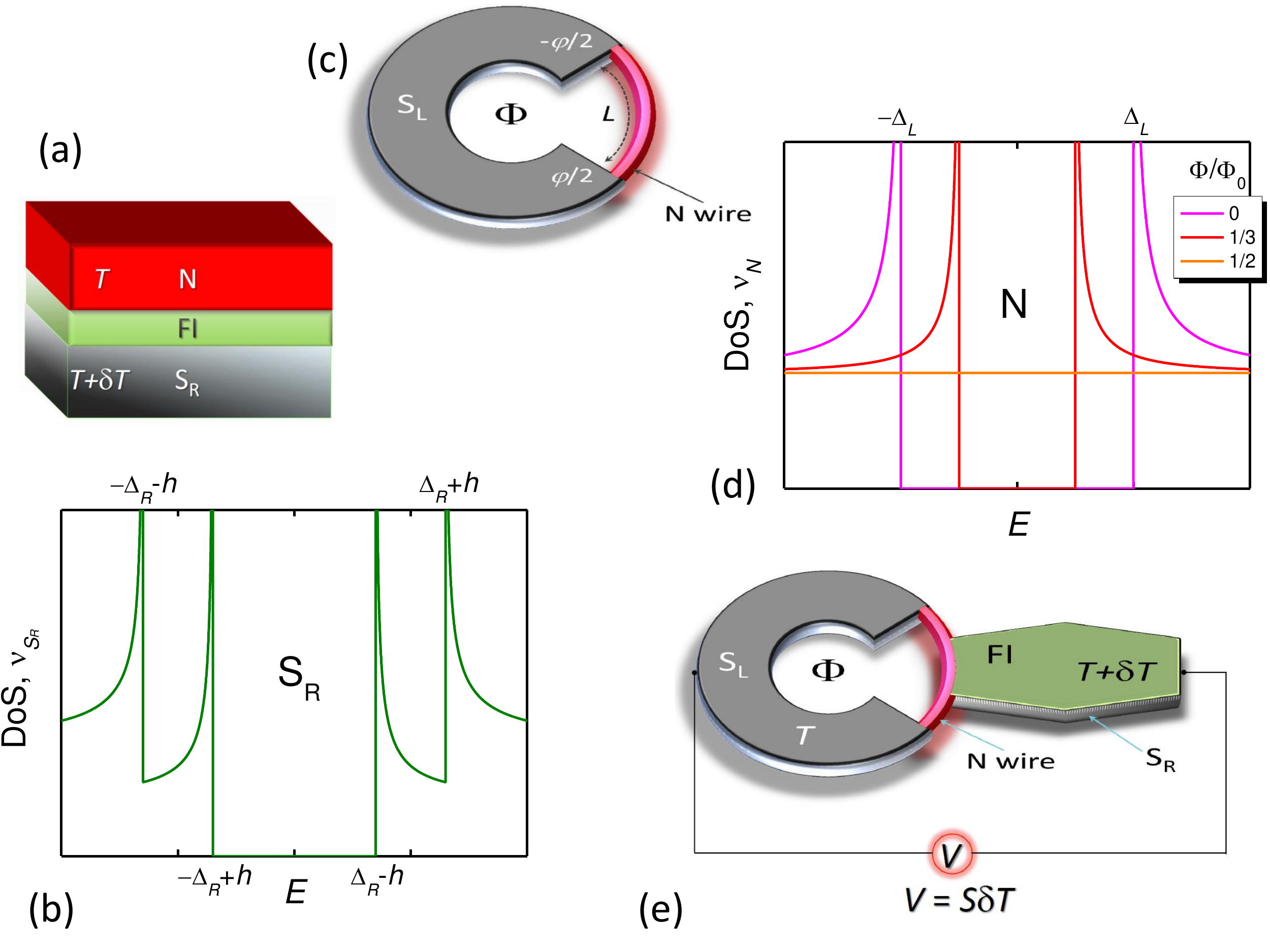}\vspace{-2mm}
\caption{\label{fig1}The phase-coherent thermoelectric transistor. (a) Sequence of stacked metallic layers consisting of a supeconductor (S$_{\text{R}}$) tunnel-coupled to a normal metal (N) through a feromagnetic insulator (FI). This implements the thermoelectric building block of the transistor. The FI layer induces an exchange field ($h$) in the superconductor and simultaneously provides a spin-polarizing barrier of polarization $P$. $T$ and $T+\delta T$ denote the temperatures of the N and S$_{\text{R}}$ elements, respectively. 
(b) Total density of states $\nu_{S_{R}}$ vs energy ($E$) in  S$_{\text{R}}$ showing the exchange-field-induced spin splitting. In the present case we set a finite $h=0.3\Delta_{R}$ where $\Delta_{R}$ is the order parameter.
(c) A superconducting quantum interference device (SQUID) containing a superconductor-normal metal-superconductor (S$_{\text{L}}$NS$_{\text{L}}$) proximity Josephson junction of length $L$. The loop is pierced by an external applied magnetic flux $\Phi$ which allows the DoS in the N region to be varied. $\varphi$ is the quantum phase difference across the N wire. The proximity SQUID implements the phase-coherent element of the transistor.
(d) Density of states $\nu_{N}$ in the N wire vs energy $E$  calculated for a few values of the applied magnetic flux. $\Phi_0$ denotes the flux quantum. 
(e)  Sketch of the resulting phase-coherent thermoelectric transistor obtained by joining the thermoelectric element in (a) with the S$_{\text{L}}$NS$_{\text{L}}$ proximity SQUID shown in (c). 
In particular, the N wire is placed on top of the FI layer, and is therefore tunnel-coupled to S$_{\text{R}}$ as in the scheme shown in  panel (a).  
Under temperature bias, a thermovoltage $V$ can develop across the transistor which can be finely tuned through the magnetic flux. $S$ indicates the transistor Seebeck coefficient.    
}
\end{center}
\end{figure*}

The phase-coherent thermoelectric transistor is based on two building blocks. The first one   is sketched in Fig. \ref{fig1}(a) and consists of  a superconducting 
film (S$_{\text{R}}$) tunnel-coupled to a normal metal (N) by  a  ferromagnetic insulator (FI). 
 The latter induces  an exchange field ($h$) in S$_{\text{R}}$ which leads  to
 a Zeeman spin-split superconducting  DoS. 
The spectrum for spin-up ($\uparrow$) and spin-down ($\downarrow$) electrons is given by 
\begin{equation}
\nu_{S_{R\uparrow(\downarrow)}}(E)=\frac{1}{2}\nu_{BCS}(E\pm h)\, ,
\end{equation}
where $\nu_{BCS}(E)=|{\rm Re}[(E+i\Gamma)/\sqrt{(E+i\Gamma)^2+\Delta_R^2(T,h)}]|$ is the conventional Bardeen-Cooper-Schrieffer DoS in a superconductor, $E$ is the energy, $\Delta_R$ is the order parameter, and $\Gamma$ accounts for broadening. 
Due to the presence of the spin-splitting field, $\Delta_R$    depends on temperature ($T$) and $h$. 
While the total DoS of S$_{\text{R}}$,  $\nu_{S_{R}}(E)=\nu_{S_{R\uparrow}}(E)+\nu_{S_{R\downarrow}}(E)$,  is electron-hole symmetric [Fig. \ref{fig1}(b)],  
 the spin-dependent $\nu_{S_{R\uparrow(\downarrow)}}(E)$ components are no longer  even functions of the energy.  
 This means that electron-hole imbalance can, in principle, be achieved using  a spin-filter contact with a normal metal.  
 This would yield a finite thermoelectric effect \cite{machon2013,ozaeta2014} in the N/FI/S$_{\textrm{R}}$ heterostructure shown in Fig. \ref{fig1}(a).  
 In addition to providing a  Zeeman splitting, the FI  also serves as a spin-filter between both metals. Therefore,  if one applies a temperature difference $\delta T$ between N and S$_{\text{R}}$, a finite charge  current ($I$) will flow through the junction due to the thermoelectric effect,  which is given by $I=\alpha\delta T/T$. 
The thermoelectric coefficient   $\alpha$ is given by \cite{ozaeta2014}
\begin{equation}
\alpha=\frac{P}{eR_T} \int_{-\infty}^\infty dE \frac{E \nu_N(E)\left[\nu_{S_{R\uparrow}}(E)-\nu_{S_{R\downarrow}}(E)\right]}{4 k_B T \cosh^2\left(\frac{E}{2k_B T}\right)}\;, \label{alpha}
\end{equation}
where  $P$ is the spin polarization due to the spin-filtering action of the FI, $T$ is the average temperature of N  and  S$_{\text{R}}$,  $k_B$ is the Boltzmann constant, $R_T$ is the junction  resistance, and $\nu_N(E)$ is the DoS of the N layer.  
%

Our proposal is based on the following observation: if one tunes the DoS in N in such a way that $\nu_N(E)$ is enhanced  at those energies where the coherent peaks of the Zeeman- split superconductor occur, then  $\alpha$ should drastically increase in  accordance to Eq. (\ref{alpha}). 
Ideally, one would search for a material  with a tunable DoS and this  can be realized  with our second building block: a  \emph{proximity} superconducting quantum interference device (SQUID), as sketched in Fig. \ref{fig1}(c). This  consists of a superconducting loop interrupted by a normal metal wire.  
The contacts between  N and  S$_{\text{L}}$ are assumed to be transparent thus allowing superconducting correlations to be induced in the N region. The latter  manifest as opening a gap in the N metal DoS with an amplitude that can be controlled via the quantum phase difference ($\varphi$) across the wire. Moreover, $\varphi$ can be controlled via an externally applied magnetic flux $\Phi$ according to $\varphi=2\pi \Phi/\Phi_0$, where $\Phi_0\simeq 2\times 10^{-15}$ Wb is the flux quantum.
If the length ($L$) of the wire is smaller than the characteristic penetration of the superconducting correlations, its DoS can be  expressed as \cite{golubov1989,heikkila2002supercurrent}
 \begin{equation}
 \label{DOSN}
 \nu_{N}(E,\Phi)=\left|{\rm Re}\left[\frac{E+i\Gamma}{\sqrt{(E+i\Gamma)^2-\Delta_L^2(T)\cos^2(\pi \Phi/\Phi_0)}}\right]\right|\; ,
\end{equation}
where 
$\Delta_L$ is the order parameter of the superconducting loop.
Equation (\ref{DOSN}) explicitly shows that $\nu_N$  can be modulated by the magnetic 
 flux. 
As displayed in Fig. 1(d),  the induced gap is fully open for vanishing flux whereas  it closes for $\Phi=\Phi_0/2$. As a consequence,  the N metal behaves as a phase-tunable superconductor. 

Our   phase-coherent thermoelectric transistor is sketched in Fig. \ref{fig1}(e). It  combines both the proximity SQUID and the S$_{\text{R}}$  spin-split superconductor.    The S$_{\text{R}}$ electrode is tunnel-coupled to the middle of   the N wire  through the FI layer.  Therefore, the  probability for electrons to tunnel  between N and S$_{\text{R}}$  depends on the sign of the spin.  The device resembles the superconducting quantum interference  proximity transistor (SQUIPT) \cite{giazotto2010squipt,meschke2011,giazotto2011,meschke2014,ronzani2014} which has been well-studied in several experiments; the crucial structural difference in our device is that the thermoelectric transistor lies in the presence of the FI layer.  
\begin{figure*}[t!]
\begin{center}
\includegraphics[width=0.75\textwidth]{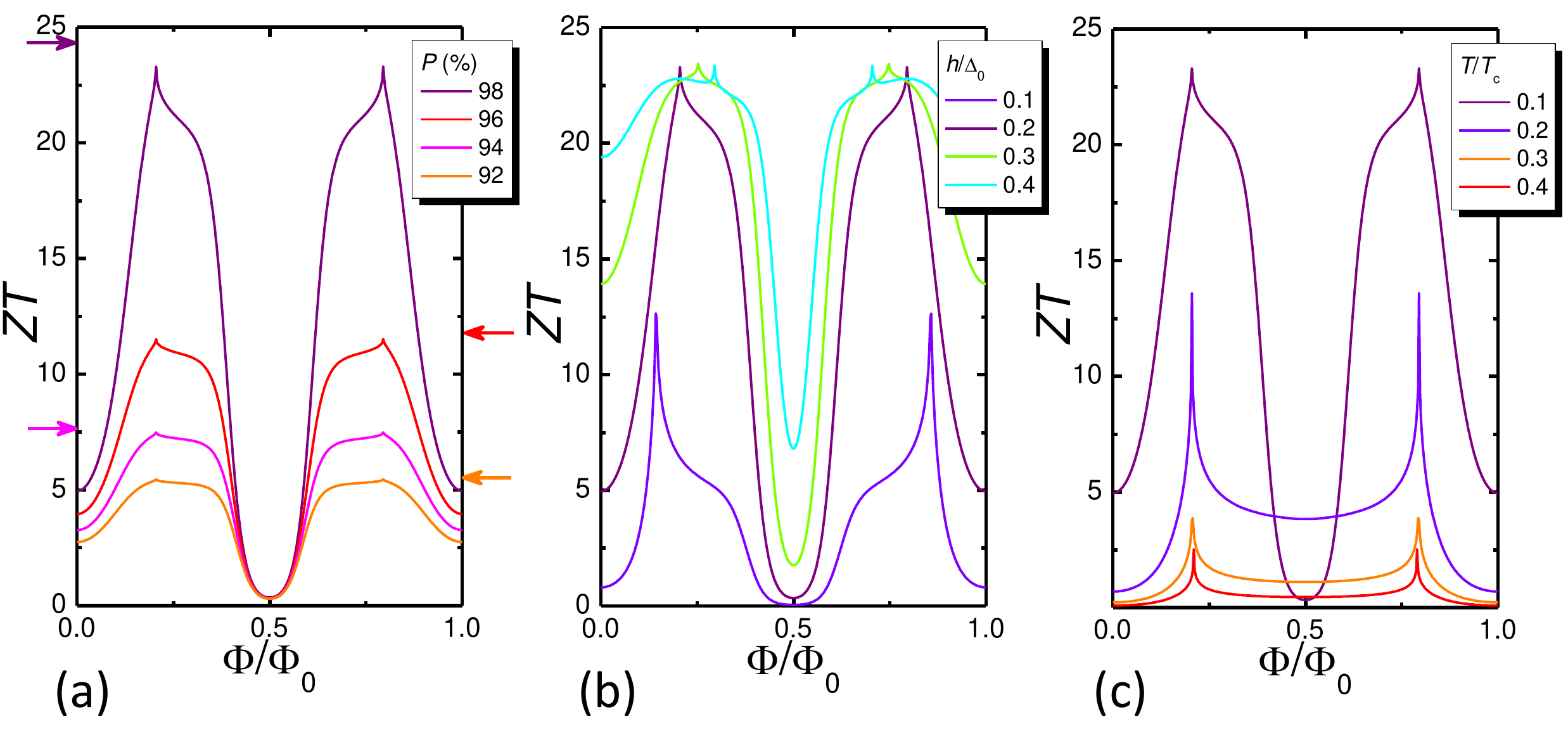}\vspace{-2mm}
\caption{Thermoelectric transistor figure of merit $ZT$. (a) Transistor themoelectric figure of merit $ZT$ vs magnetic flux $\Phi$ calculated for a few polarization  values ($P$) of the FI layer at $h=0.2\Delta_0$ and $T=0.1 T_c$. Colored arrows indicate the zero-temperature limit $ZT=P^2/(1-P^2)$ for the given barrier polarizations.
(b) $ZT$ vs $\Phi$ calculated for some exchange field values at $T=0.1 T_c$ and $P=98\%$. 
(c) $ZT$ vs $\Phi$ calculated for a few temperatures $T$ assuming $P=98\%$ and $h=0.2\Delta_0$. $T_c\approx \Delta_0/1.764 k_B$ is the common critical temperature of both superconductors forming the thermoelectric device, and $\Delta_0$ is the zero-temperature, zero-exchange field energy gap. 
}
\label{fig2}
\end{center}
\end{figure*}

The efficiency of the  thermoelectric transistor  can be  finely tuned by the applied the magnetic flux. 
To determine the ideal conditions for large thermoelectric efficiency we calculate the linear response transport coefficients. In other words, we calculate the charge ($I$) and heat ($\dot Q$)  currents flowing  through the structure  when either a small voltage $V$ or a small temperature difference $\delta T$ is applied across the junction. At steady state these currents are  expressed as follows: 
\begin{equation}
 \begin{pmatrix} I \\ \dot Q  \end{pmatrix} = \begin{pmatrix} \sigma &  \alpha  \\ \alpha & \kappa T  \end{pmatrix} \begin{pmatrix} V \\ \delta T/T \end{pmatrix}\; .\label{response1}
\end{equation}
Here, $\alpha$ is the thermoelectric coefficient defined in Eq. (\ref{alpha}) which has now become phase dependent through  $\nu_N(E,\Phi)$.  As well as the  the electric  ($\sigma$)  and  the thermal  ($\kappa$) conductances given by the expressions: \cite{ozaeta2014}
\begin{subequations}
  \label{eq:numcoefs}
  \begin{align}
    \sigma(\Phi)&=\frac{1}{R_T} \int_{-\infty}^\infty dE \frac{\nu_N(E,\Phi)\left[\nu_{S_{R\uparrow}}(E)+\nu_{S_{R\downarrow}}(E)\right]}{4 k_B T \cosh^2\left(\frac{E}{2k_B T}\right)}
    \,,
    \\
    \kappa(\Phi)&=\frac{1}{e^2R_T} \int_{-\infty}^\infty dE \frac{E^2 \nu_N(E,\Phi)\left[\nu_{S_{R\uparrow}}(E)+\nu_{S_{R\downarrow}}(E)\right]}{4 k_B T^2 \cosh^2\left(\frac{E}{2k_B T}\right)}
    \,.
\end{align}
\end{subequations}

The thermoelectric efficiency of the transistor 
 can be quantified by the usual dimensionless figure of merit $ZT$  defined as
\begin{equation}
ZT=\frac{S^2\sigma T}{\tilde\kappa}\; ,\
\end{equation}
where $\tilde\kappa=\kappa-\alpha^2/\sigma T$ is the thermal conductance at zero current and  $S=-\alpha/(\sigma T)$ the Seebeck coefficient that is a measure  for  the thermopower [see Fig. \ref{fig1}(e)].  
We stress that there is no theoretical limit for $ZT$: if it approaches infinity the efficiency of the transistor would reach the Carnot limit. 
So far, only exceptionally-good  thermoelectric bulk materials are able to provide values of $ZT$ slightly larger than 1 [see Fig. \ref{fig4}(a)]. 
By contrast, our thermoelectric transistor can result in $ZT$ values as large as  several tens thanks to the fine tuning offered via the phase. 
This is displayed in Fig. \ref{fig2} where the flux dependence of $ZT$ as a function of different parameters is shown. 
In Fig. \ref{fig2}(a) the temperature and splitting field are fixed at moderate values,  $0.1 T_c$ and $0.2\Delta_0$, respectively. \cite{Meservey1994,moodera2007phenomena,hao1990spin,Moodera1993} $T_c \approx (1.764k_B)^{-1}\Delta_0$ denotes the superconducting critical temperature, $\Delta_0$ is the zero-temperature, zero-exchange field energy gap, and 
for clarity we assume that both S$_{\text{L}}$ and S$_{\text{R}}$ have the same $\Delta_0$ value.  
\begin{figure}[b!]
\includegraphics[width=\columnwidth]{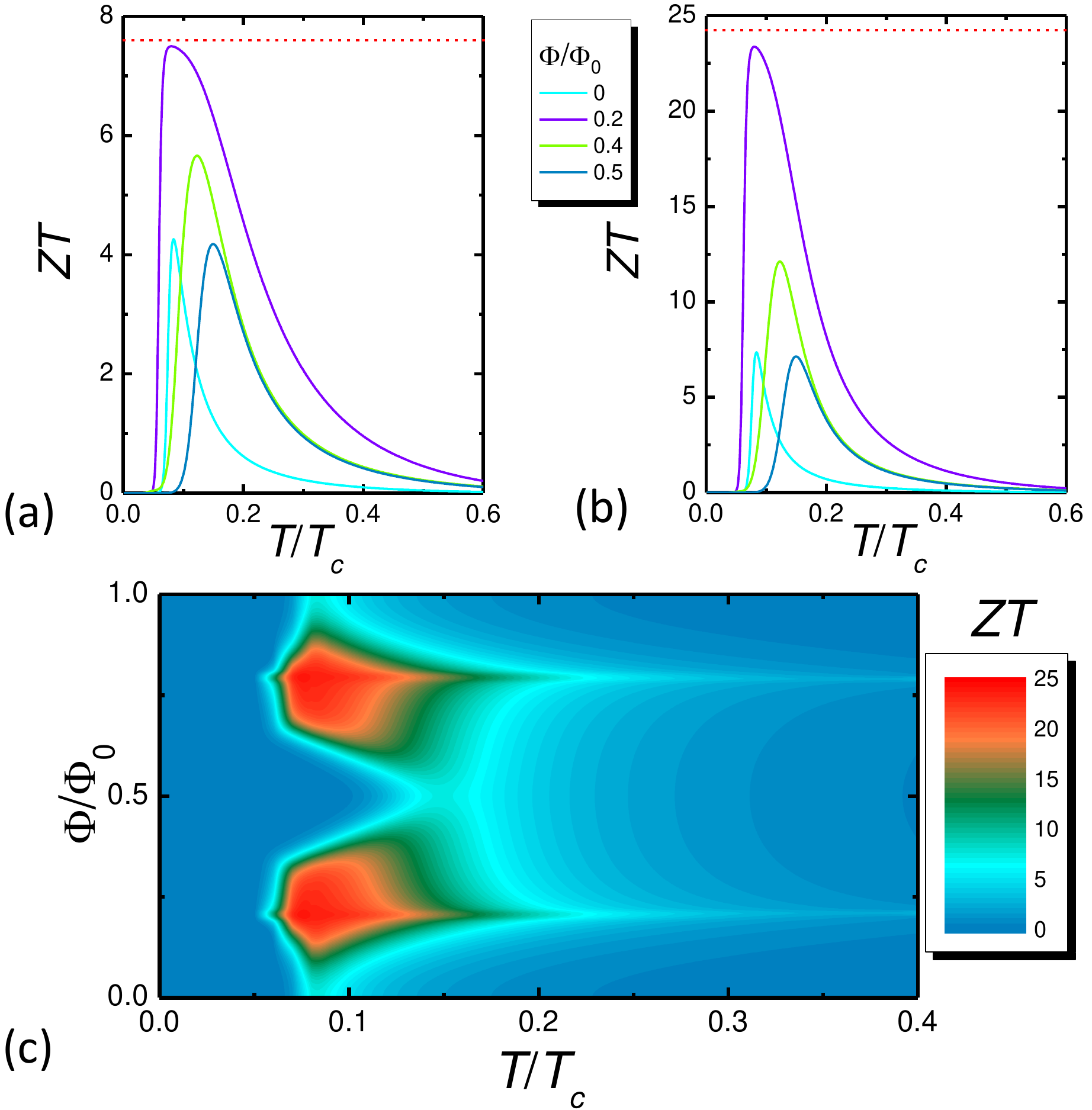}\vspace{-2mm}
\caption{Full temperature and magnetic flux behavior of the thermoelectric figure of merit. (a) Thermoelectric figure of merit $ZT$ vs temperature $T$ calculated for a few values of the applied flux $\Phi$ at $h=0.2\Delta_0$ and $P=94\%$. (b) The same as in panel (a) but calculated for $P=98\%$. Red dashed lines in panels (a) and (b) indicate the zero-temperature limit  for the above given barrier polarizations. 
(c) Color plot  of $ZT$ vs $T$ and $\Phi$ calculated at $h=0.2 \Delta_0$ and for $P=98\%$.
}
\label{fig3}
\end{figure}
\begin{figure*}[t!]
\includegraphics[width=0.75\textwidth]{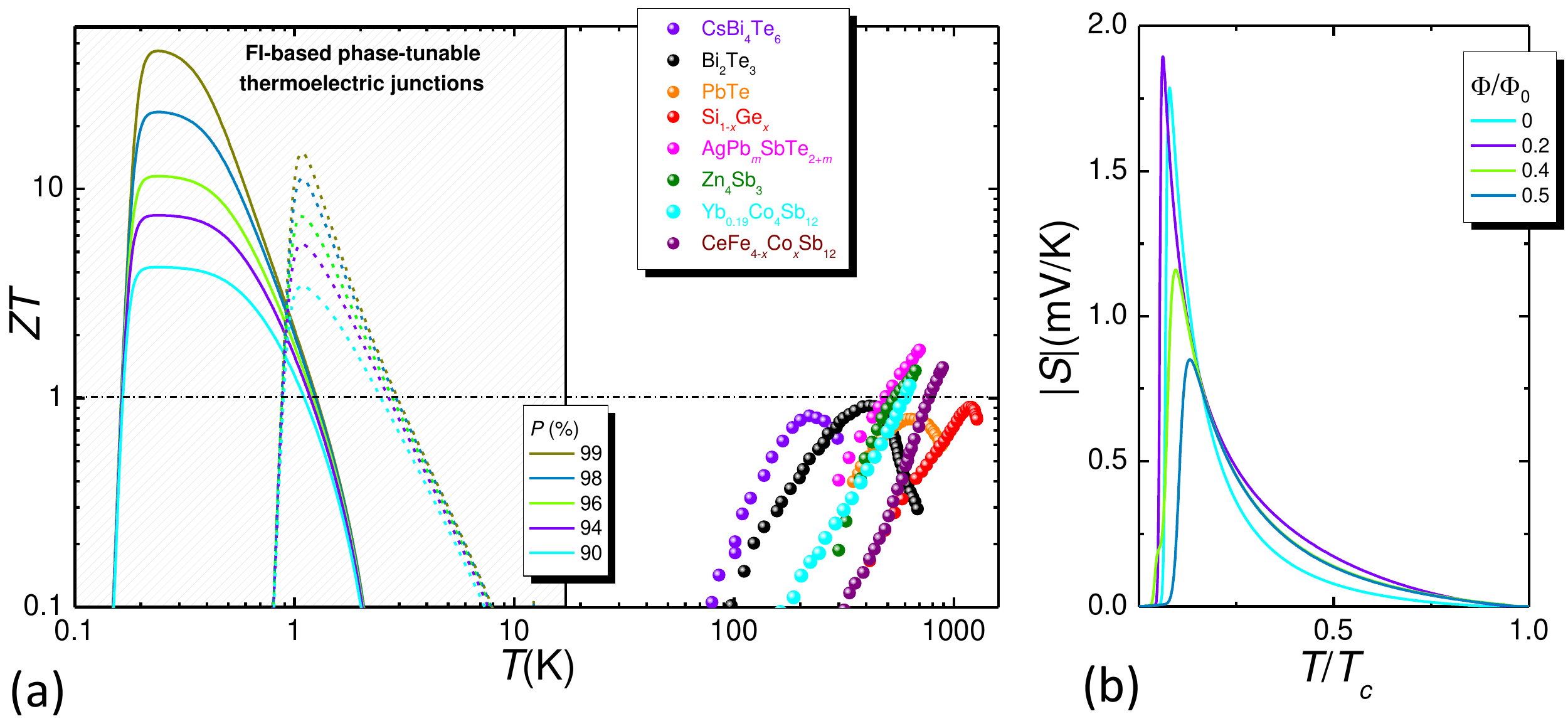}\vspace{-2mm}
\caption{Comparison of the transistor performance with state-of-the-art high-$ZT$ materials and the achievable Seebeck coefficient. (a) Left side:  Expected figure of merit $ZT$ vs $T$ for an EuS/Al-type (solid lines) and GdN/NbN-type (dashed lines) phase-coherent thermoelectric transistors calculated for selected values of the barrier polarization.  We assumed $T_c=3$K and $h=0.2\Delta_0$ for the former structure whereas for the latter we set $T_c=14$K  and $h=0.1\Delta_0$. Furthermore, for both of them we set $\Phi=0.2\Phi_0$.
Right side:  Figure of merit $ZT$ vs temperature possessed by several state-of-the-art commercial bulk thermoelectric materials. 
 Dash-dotted line indicates $ZT=1$.
 (b) Absolute value of the Seebeck coefficient ($S$) achievable in the phase-coherent thermoelectric transistor vs $T$ calculated for a few selected values of the applied magnetic flux at $h=0.2\Delta_0$ and $P=98\%$.  Data shown in panel (a) are taken from Ref.\cite{tritt2006thermoelectric}.
}
\label{fig4}
\end{figure*}
Therefore,  by tuning the flux through the loop the theoretical zero-temperature limit assuming no inelastic scattering processes for $ZT=P^2/(1-P^2)$ (indicated by  colored arrows corresponding to each $P$ value) \cite{ozaeta2014} can be approached without requiring ultra-low temperatures. It is also apparent  that a high barrier polarization is crucial in order to achieve large values of $ZT$.  Good candidates for the tunneling barriers are  europium chalcogenides (EuO and EuS) for which values of $P$ ranging from 80 up to almost 100\%  have been reported. \cite{Moodera:1988bb,hao1990spin,Moodera1993,santos2004observation,moodera2007phenomena,Santos:2008cm,miao2009controlling,li2013superconducting}  Also very high spin-filtering has been reported in   GdN barriers \cite{Senapati:2011fm,PalAdvanced,pal2014pure} with polarizations as large as $97\%$ at 15 K (even larger values are expected as $T$ decreases).  

In Fig. \ref{fig2}(b) $P$ is set to 98\%,  $T=0.1T_c$ and $ZT$ for different values of the induced Zeeman field $h$ is plotted showing an interesting 
 non-monotonic behavior.
$h$  cannot be externally tuned but can be partially controlled during growth  of the FI/S$_{\textrm{R}}$ interface since it depends on the quality of the contact. Values from 0.1 meV up to few meV  for the Zeeman splitting have been reported. \cite{Tedrow1986,Hao1991,moodera2007phenomena,li2013superconducting}  

We emphasize that the amplitude of the thermoelectric effect also depends on the length of the N wire. We assumed that $L$ is smaller than the superconducting coherence length  [see Eq. (\ref{DOSN})] or meaning $\Delta_0<\epsilon_{Th}$, where $\epsilon_{Th}=\hbar D/L^2$ is the Thouless energy and $D$ the diffusion coefficient of the N wire.  Although these two energies in the  existing experiments on   SQUIPTs are of same order \cite{meschke2014,ronzani2014}, the observed modulation of the induced gap by the magnetic flux can be well described with  Eq. (\ref{DOSN}) \cite{ronzani2014}.  Instead of the N bridge one can use a short constriction made with the same material of the loop.   In such a case one avoids  the mismatch of the Fermi velocities at the interfaces, meaning the proximity effect between the bulky part of the loop and the constriction is increased.

Figure \ref{fig2}(c) shows the flux dependency of $ZT$  for different values of the temperature,  at $P=98\%$ and $h=0.2\Delta_0$. It is clear that the efficiency of the transistor  decreases by increasing the temperature towards $T_c$.

 Figures \ref{fig3}(a) and \ref{fig3}(b) show the temperature dependence of $ZT$ for different magnetic fluxes. $ZT$  reaches a maximum  at a finite $T$  that corresponds to the temperature  for   which  the difference between the gap induced in N and the gap in the S$_{\text{R}}$ electrode matches the value of the exchange field, i.e., when the coherent peaks in the DoS at the edges of the gaps [see Figs. \ref{fig1}(b) and \ref{fig1}(d)]  coincide.  
 The red dashed lines in Figs. \ref{fig3}(a) and \ref{fig3}(b) show the upper theoretical  value of $ZT$ at zero temperature \cite{ozaeta2014}.  Our device, however,  allows to approach this maximum value  at finite $T$ by tuning the magnetic flux.
The full dependence of $ZT$ on both $T$ and $\Phi$, for $P=98\%$ and $h=0.2\Delta_0$ is shown by 
the color plot in  Fig. \ref{fig3}(c).  For $T\approx 0.1T_c$ and $\Phi\approx0.25 \Phi_0$ or $\Phi\approx0.75 \Phi_0$ the figure of merit can reach values close to $P^2/(1-P^2)\approx 25$.

To place our device in the context of thermoelectrics we show in Fig. \ref{fig4}(a) the state-of-the-art bulk materials with the highest $ZT$ values.  One of the most widely used material is 
Bi$_2$Te$_3$, \cite{tritt2006thermoelectric,snyder2008} which at room temperature shows a $ZT$ close to 1. All other materials show their highest values of $ZT$ at high temperatures. By contrast, our transistor operates at low temperatures where it can reach values of $ZT$ which are larger by more than an order of magnitude.  An appropriate candidate is  EuO combined with superconducting Al where we expect a $ZT$ of $\sim 30-40$ for realistic values of the spin-filter efficiency. \cite{santos2004observation,Santos:2008cm} Other  suitable Eu chalcogenides include EuS or EuSe. In the latter case $P$ and $h$ can be tuned by an external magnetic field. \cite{Moodera1993} An alternative to the chalcogenides  is to use  GdN with superconducting NbN. \cite{Senapati:2011fm,pal2014pure} Its advantage is the higher critical  temperature of NbN of $\simeq 15$ K. 
As for the normal metal bridge, it is important  to achieve good electric contact  with the S$_{\text{R}}$ loop in order to develop a sizeable and tunable proximity gap. If, for instance, one uses Al as superconductor for the ring,  suitable candidates for the N wire are copper (Cu) \cite{giazotto2010squipt,meschke2011,meschke2014,ronzani2014} or silver (Ag) \cite{sns2008}.
 
 The temperature-voltage conversion capability of the thermoelectric transistor can be quantified by the Seebeck coefficient $S$: its temperature dependence is shown in Fig. \ref{fig4}(b) for selected values of the applied magnetic flux.  Here we set $P=98\%$ and $h=0.2\Delta_0$.
 The transistor is extremely sensitive at low temperatures, and provides a sharp response even to tiny temperature gradients. In particular,  Seebeck coefficients as large as a few mV/K can be achieved under optimal flux tuning conditions, which have to be compared to coefficients as large as a few hundreds of $\mu$V/K obtained in the abovementioned  high-performance thermoelectric materials.
These sizeable values of $S$ make our phase-coherent thermoelectric transistor an ideal candidate for the implementation of  cryogenic power generators or ultrasensitive low-temperature general purpose thermometry as well as, for  radiation sensors  where  heating of one of the superconductors forming the junction is achieved due to a coupling with radiation. 
 
The work of F.G. 
has been  partially funded by
the European Research Council under the European Union's Seventh Framework Programme (FP7/2007-2013)/ERC grant agreement No. 615187-COMANCHE, and by
the Marie Curie Initial Training Action (ITN) Q-NET 264034. 
J.W.A.R. acknowledges funding from the Royal Society ("Superconducting Spintronics"). J.W.A.R. and F.S.B. jointly acknowledge funding from the Leverhulme Trust through an International Network Grant (grant IN-2013-033). 
J.S.M acknowledges funding from the National Science Foundation ((grant DMR 1207469) and the Office of Naval Research (grant N00014-13-1-0301).
The work of F.S.B. has been
supported by the Spanish Ministry of Economy and Competitiveness under
Project No. FIS2011-28851-C02-02.

%
%


\end{document}